\documentclass[runningheads]{llncs}

\usepackage{times}
\usepackage{helvet}
\usepackage{courier}
\usepackage[utf8]{inputenc}

\usepackage{mymacros}

\begin{document}
\setlength{\marginparwidth}{3cm}

\title{Adversarial Deep Reinforcement Learning based Adaptive Moving Target Defense}

\author{Taha Eghtesad\inst{1} \and
Yevgeniy Vorobeychik\inst{2} \and
Aron Laszka\inst{1}}

\authorrunning{T. Eghtesad et al.}

\institute{University of Houston, Houston, TX 77054, USA \and
Washington University in St. Louis, St. Louis, MO, 63130}

\maketitle


\begin{abstract}
Moving target defense (MTD) is a proactive defense approach that aims to thwart attacks by continuously changing the attack surface of a system (e.g., changing host or network configurations), thereby increasing the adversary's uncertainty and attack cost. To maximize the impact of MTD, a defender must strategically choose when and what changes to make, taking into account both the characteristics of its system as well as the adversary's observed activities.
Finding an optimal strategy for MTD presents a significant challenge, especially when facing a resourceful and determined adversary who may respond to the defender's actions. In this paper, we propose a multi-agent partially-observable Markov Decision Process model of MTD and formulate a two-player general-sum game between the adversary and the defender.
Based on an established model of adaptive MTD, we propose a multi-agent reinforcement learning framework based on the double oracle algorithm to solve the game. In the experiments, we show the effectiveness of our framework in finding optimal policies.
\end{abstract}

\section{Introduction}
\label{sec:intro}

Traditional approaches for security focus on preventing intrusions (e.g, hardening systems to decrease the occurrence and impact of vulnerabilities) or on detecting and responding to intrusions (e.g., restoring the configuration of  compromised servers). While these passive and reactive approaches are useful, they cannot provide perfect security in practice.
Further, these approaches let adversaries perform reconnaissance and planning unhindered, giving them a significant advantage in information and initiative. 
As adversaries are becoming more sophisticated and resourceful, it is imperative for defenders to augment traditional approaches with more proactive ones, which can give defenders the upper hand. 

\textit{Moving Target Defence} (MTD) is a proactive approach that changes the rules of the game in favor of the defenders. MTD techniques enable defenders to thwart cyber-attacks by continuously and randomly changing the configuration of their assets (i.e., networks, hosts, etc.). These changes increase the uncertainty and complexity of attacks, making them computationally expensive for the adversary \cite{zheng2019markov} or putting the adversary in an infinite loop of exploration~\cite{tan2019optimal}.


Currently, system administrators typically have to manually select MTD configurations to be deployed on their networked systems based on their previous experiences~\cite{hu2019reinforcement}. This has two main limitations. First, it can be very time consuming since there are constraints on data locations, physical connectivity of servers cannot be easily changed, and resources are limited. Second, it is difficult to capture the trade-off between security and efficiency~\cite{chen2015practical}.

In light of this, it is crucial to provide automated approaches for deploying MTDs, which maximize security benefits for the protected assets. This requires a design model that reflects multiple aspects of the MTD environment \cite{li2019optimal,zheng2019markov,prakash2015empirical,albanese2019moving}. Further, we need a decision making algorithm for the model to select which technique to deploy and where to deploy it \cite{tan2019optimal}. Finding optimal strategies for deployment of MTDs are computationally challenging. For example, the adversary can adapt to MTD deployments, or the state-action space of the environment can be huge even for trivial number of MTD configurations or in-control assets.



Recently, many research efforts have applied \textit{Independent Reinforcement Learning} (InRL) techniques to find the best action policies in known or unknown decision making environments, such as cybersecurity. In InRL, an agent learns to make the best decision by continuously interacting with its unknown environment. In general, traditional reinforcement learning techniques use tabular approaches to store estimated rewards (e.g., $Q$-Learning)~\cite{iannucci2019performance}. To address challenges of reinforcement learning such as exploding state-action space, \textit{Artificial Neural Networks} (ANN) have replaced table based approaches in many domains, thereby decreasing the training time and memory requirements. This led to the emergence of \textit{deep reinforcement learning} (DRL) algorithms such as D$Q$L~\cite{mnih2013playing}.

The naive approach to multi-agent reinforcement learning (MARL) is to use InRL where one player treats the opponent's action as part of its localized environment~\cite{eghtesad2019deep}. However, two problems arise here: 1) convergence guarantees are lost since these localize environments are non-stationary and non-Markovian~\cite{laurent2011world}; and 2) These policies can not generalize well since they overfit to the opponent’s policies~\cite{lanctot2017unified}.

\paragraph{Contributions}
We create a multi-agent partially-observable Markov decision process for MTD, and based on this model we propose a two-player general-sum game between the adversary and the defender. We present a multi-agent deep reinforcement learning approach to solve the game. Our main contributions are as follows:
\begin{itemize}
    \item We propose a multi-agent partially-observable Markov decision process for MTD.
    \item We propose a two-player general-sum game between the adversary and the defender based on this model.
    \item We formulate the problem of finding adaptive MTD policies as finding the mixed strategy Nash equilibrium of this game.
    \item We propose a compact memory representation for the defender and adversary agents, which helps them to better operate in the partially observable environment.
    \item We propose a computational approach for finding the optimal MTD policy using Deep $Q$-Learning and the Double Oracle algorithms.
    \item We evaluate our approach while exploring various game parameters.
    \item We show that our approach is viable in terms of computational cost.
\end{itemize}

\paragraph{Organization}
The rest of the paper is organized as follows.
In Section~\ref{sec:model}, we introduce a multi-agent partially-observable Markov decision process for MTD which is used as the basis of the MARL.
In Section~\ref{sec:preliminaries}, we describe the preliminaries such as the InRL (Section~\ref{subsec:rl}), and one of InRL family algorithms, \ie Deep $Q$ Learning (Section~\ref{subsec:dql}).
In Section~\ref{sec:problem}, we formulate a two-player general-sum game between the adversary and the defender, and formulate the problem of finding adaptive MTD policies as finding the MSNE of the game.
In Section~\ref{sec:framework}, we discuss the solution overview (Section~\ref{subsec:do}), discuss the challenges faced when solving the game (Section~\ref{subsec:challenges}), and propose our framework to solve this game (Section~\ref{subsec:approach}). 
In Section~\ref{sec:eval}, we provide a thorough numerical analysis of our approach.
In Section~\ref{sec:related}, we discuss the related work.
Finally, in Section~\ref{sec:conclusion}, we provide concluding remarks and outline of directions for future~work.

\section{Model}
\label{sec:model}

To model adaptive \emph{Moving Target Defense}, we build a \emph{Multi-Agent Partially-Observable Markov Decision Process} (MAPOMDP) based on the model of Prakash and Wellman~\cite{prakash2015empirical}. In this adversarial model, there are two players, a defender and an adversary, who compete for control over a set of servers.
At the beginning of the game, all servers are under the control of the defender.
To take control of a server, the adversary can launch a \emph{``probe''} against the server at any time, which either compromises the server or increases the success probability of subsequent probes.
To keep the servers safe, the defender can \emph{``reimage''} a server at any time, which takes the server offline for some time, but cancels the adversary's progress and control.
The goal of the defender is to keep servers uncompromised (i.e., under the defender's control) and available (i.e., online).
The goal of the adversary is to compromise the servers or make them unavailable. 
For a list of symbols used in this paper, see Table~\ref{tab:symbols}.

\begin{table}[t]
    \centering
    \caption{List of Symbols and Experimental Values}
    \label{tab:symbols}
        \begin{tabular}{|c|p{9.0cm}|c|}
            \hline
            Symbol & Description & Baseline Value \\
            \hline
            \multicolumn{3}{|c|}{Environment, Agents, Actions} \\
            \hline
            \rowcolor{Gray} $M$ & number of servers & 10 \\
            $\Delta$ & number of time steps for which a server is unavailable after reimaging & 7\\
            \rowcolor{Gray} $\nu$ & probability of the defender not observing a probe & 0 \\
            $\alpha$ & knowledge gain of each probe & 0.05 \\
        \rowcolor{Gray} $C_A$ & attack (\textit{probe}) cost & 0.20\\
            $\theta_{sl}^p$ & slope of reward function & 5 \\
            \rowcolor{Gray} $\theta_{th}^p$ & steep point threshold of reward function & 0.2 \\
            $w^p$ & weighting of reward for having servers up and in control of adversary / defender & 0 / 1 \\
            \rowcolor{Gray} $r^p_\tau$ & reward of player $p$ in time step $\tau$ & \\
            \hline
            
            \multicolumn{3}{|c|}{Heuristic Strategies} \\
            \hline
            \rowcolor{Gray} $P_D$ & period for defender's periodic strategies & 4 \\
            $P_A$ & period for adversary's periodic strategies & 1 \\
            \rowcolor{Gray} $\pi$ & threshold of number of probes on a server for \emph{PCP} defender & 7 \\
            $\tau$ & threshold for adversary's / defender's \emph{Control-Threshold} strategy & 0.5 / 0.8 \\
            \hline
            
            \multicolumn{3}{|c|}{Reinforcement Learning} \\
            \hline
            $T$ & length of the game (number of time steps) & 1000 \\
            \rowcolor{Gray} $\gamma$ & temporal discount factor & $0.99$ \\
            $\epsilon_p$ & exploration fraction & $0.2$ \\
            \rowcolor{Gray} $\epsilon_f$ & final exploration value & $0.02$ \\
            $\alpha_t$ & learning rate & $0.0005$ \\
            \rowcolor{Gray} $\vert E \vert$ & experience replay buffer size & $5000$ \\
            $\vert X \vert$ & training batch size & $32$ \\
            \rowcolor{Gray} $N_e$ & number of training episodes &  $500$\\
            \hline
        \end{tabular}
\end{table}

\subsection{Environment and Players}
\label{subsec:env}
There are $M$ \emph{servers} and two players, a \emph{defender} and an \emph{adversary}.
The servers are independent of each other in the sense that they are independently attacked, defended, and controlled. The game environment is explained in detail in the following subsections.

\subsection{State}
\label{subsec:state}

Time is discrete, and in a given time step $\tau$, the state of each server $i$ is defined by tuple
    $s_i^\tau = \langle \rho, \chi, \upsilon \rangle$
where 
\begin{itemize}
    \item \textbf{$\rho \in \mathbb{Z}^*$} represents the number of probes lunched against server $i$ since the last reimage,
    
    \item \textbf{$\chi \in \{\field{adv},\field{def}\}$} represents the player controlling the server, and 
    
    \item \textbf{$\upsilon \in \{\field{up}\}\cup\mathbb{Z}^*$} represents if the server is online (i.e., up) or if it is offline (i.e., down) with the time step in which the server was reimaged.
\end{itemize}

\subsection{Actions}
\label{subsec:actions}
In each time step, a player may take either a single action or no action at all.
The adversary's action is to select a server and \emph{probe} it. 
Probing a server takes control of it with probability
\begin{align}
    1-e^{-\alpha\cdot(\rho+1)}
\end{align}
where $\rho$ is the number of previous probes and $\alpha$ is a constant that determines how fast the probability of compromise grows with each additional probe, which captures how much information (or progress) the adversary gains from each probe. Also, by probing a server, the adversary can understand whether it is up or down.

The defender's action is to select a server and \emph{reimage} it.
Reimaging a server takes the server offline for a fixed number $\Delta$ of time steps, after which the server goes online under the control of the defender and with the adversary's progress (i.e., number of previous probes $\rho$) against that server erased (i.e., reset to zero).

\subsection{Rewards}
\label{subsec:rewards}
Prakash and Wellman~\cite{prakash2015empirical} define a family of utility functions. 
The exact utility function can be chosen by setting the values of preference parameters, which specify the goal of each player. 
The value of player $p$'s utility function $u^p$, as described by Equation~\ref{eq:reward} and Equation~\ref{eq:reward_sigmoid}, depends on the number of servers in control of player $p$ and the number of servers offline. Note that the exact relation depends on the scenario (e.g., whether the primary goal is confidentiality or integrity), but in general, a higher number of in control servers yields a higher utility.

\begin{align}
\label{eq:reward}
    u^p(n_c^p, n_d) = w^p \cdot f\left(\frac{n_c^p}{M},\theta^p\right) + (1-w^p) \cdot f\left(\frac{n_c^p + n_d}{M},\theta^p\right)
\end{align}
where $n_c^p$ is the number of servers which are up and in control of player $p$, $n_d$ is the number of unavailable (down) servers, and $f$ is a sigmoid function with parameters $\theta$:
\begin{align}
\label{eq:reward_sigmoid}
    f(x,\theta) = \frac{1}{\mathrm{e}^{-\theta_{sl}\cdot (x-\theta_{th})}}
\end{align}
where $\theta_{sl}$ and $\theta_{th}$ control the slope and position of the sigmoid's steep point, respectively. 

\begin{wraptable}{r}{.4\linewidth}
    \centering
        \caption{Utility Environments}
    \label{tab:util_env}
        \begin{tabular}{|c|c|c|c|}
            \hline
            & \textbf{Utility Environment} & $w^a$  & $w^d$ \\ \hline
            0 & control / availability & 1 & 1 \\
            \rowcolor{Gray} 1 & control / confidentiality & 1 & 0 \\
            2 & disrupt / availability & 0 & 1 \\
            \rowcolor{Gray} 3 & disrupt / confidentiality & 0 & 0 \\
            \hline
        \end{tabular}

\end{wraptable}

Reward weight
($w^p$) specifies the goal of each player. As described by Prakash and Wellman~\cite{prakash2015empirical}, there can be four extreme combinations of this parameter, which are summarized in Table~\ref{tab:util_env}.
For example, in \textit{control / availability}, both players gain reward by having the servers up and in control. Or in \textit{disrupt / availability}, which is the most interesting case, the defender gains reward by having the servers up and in control, while the adversary gains reward by bringing down the servers or having them in control.

The defender's cost of action is implicitly defined by the utility function. In other words, cost of reimaging a server comes from not getting reward for the times that it is ``down.'' However, the adversary's reward accounts for the cost of probing ($C_A$), which is a fixed costs that can be avoided by not taking any action.

The reward given to the adversary ($r^a_\tau$) and defender ($r^d_\tau$) at time $\tau$ is defined by:
\begin{align}
    r^a_\tau &= \begin{cases}
            u^a(n_c^a, n_d) - C_A & \text{adversary probed a server at } \tau \\
            u^a(n_c^a, n_d) & \text{adversary did nothing}
        \end{cases} \\
    r^d_\tau &= u^d_\tau
\end{align}

\subsection{Observations}
\label{subsec:obs}


A key aspect of the model is the players' uncertainty regarding their state.
The defender does not know which servers have been compromised by the adversary. Also, the defender observes each probe with a fixed probability $1 - \nu$ (with probability $\nu$, the probe is undetected).
Consequently, the defender can only estimate the number of probes against a server and whether a server is compromised.
However, the defender knows the the status of all servers (i.e., whether the server is up or down, or if it is down, how many time steps it requires to be back up again).

The adversary always observes when the defender reimages a compromised server, but cannot observe reimaging an uncompromised server without probing it.
Consequently, the adversary knows with certainty which servers are compromised. 

Observation of a player $p$ at time $\tau$ is defined as a vector of tuples $O_i^p$ where $O_i^p$ corresponds to observation of $p$ of server $i$.
\begin{align}
    &O^p_\tau = \langle O_{1, \tau}^p, O_{2, \tau}^p, \cdots, O_{M, \tau}^p \rangle
\end{align}{}

The adversary knows which servers are compromised and knows how many attacks it has initiated on each server. Also, if the server is down, the adversary can estimate the time that the server is up again. The observation of a server $i$ for adversary at time $\tau$ is defined as a tuple $O_{i, \tau}^a$:
\begin{align}
    \forall_{0 \leq i < M}: && O_{i, \tau}^a = \langle \field{status}, \field{time\_to\_up}, \field{progress}, \field{control} \rangle 
\end{align}
where $\field{status} \in \{1, 0\}$ which shows the server is up or down, and $\field{control} \in \{1, 0\}$ shows that the adversary controls that server or not, respectively.

Observation vector of the defender is almost the same as the adversary. The only difference is that the defender does not know who controls the servers, and only has a estimation on the number of probes (where $\nu$ is not $0$):
\begin{align}
    \forall_{0 \leq i < M}: && &O_{i, \tau}^d = \langle \field{status}, \field{time\_to\_up}, \field{progress} \rangle 
\end{align}

To overcome challenges regarding the defender which are described in Section~\ref{subsec:challenges}, we included two more attributes to have some form of memory for the defender. These two attributes are described in Section~\ref{subsec:partial}.

\section{Preliminaries}
\label{sec:preliminaries}

In this section, we describe the family of reinforcement learning algorithms (Section~\ref{subsec:rl}), and one algorithm in this family, namely the Deep $Q$-Learning (Section~\ref{subsec:dql}).

\subsection{Independent Reinforcement Learning}

\label{subsec:rl}

One of the main approaches for finding a decision making policy is the \emph{Independent Reinforcement Learning} (InRL) which focuses on interactions of a single agent and the environment, in order to maximize the agent's gain (presented as rewards or utilities) from the environment. Figure~\ref{fig:inrl} shows the interactions between different components of InRL. Further, the InRL algorithm is described in Algorithm~\ref{algo:inrl}. A basic InRL environment is a \textit{Partially-Observable Markov Decision Process} (POMDP), which can be represented as a tuple:
\begin{align}
\label{eq:mdp}
    \field{POMDP} = \langle \mathbb{S}, \mathbb{A}, \mathbb{T}, R, \mathbb{O} \rangle.
\end{align}
where $\mathbb{S}$ is the set of all possible states in the environment (described in Section~\ref{subsec:state}), $\mathbb{A}$ is the set of all possible actions by the agent, $\mathbb{T}$ is the set of stochastic transition rules (Section~\ref{subsec:state} and Section~\ref{subsec:actions}), $R$ is the immediate reward of a state transition (Section~\ref{subsec:rewards}), and $\mathbb{O}$ is the set of observation rules of the agent (Section~\ref{subsec:obs}).

\begin{wrapfigure}{r}{.4\textwidth}
    \centering
    \resizebox{\linewidth}{!}{
        \tikzstyle{block} = [draw, fill=gray!20, rectangle,
            minimum height=2em, minimum width=4em]
            
        \tikzstyle{line} = [draw, -latex]
        
        \begin{tikzpicture}[node distance = 6em, auto]
            \node [block] (Agent) {Agent};
            \node [block, below of=Agent] (Environment) {Environment};
            
                 \path [line] (Agent.0) --++ (4em,0em) |- node [anchor=center, above, near start, rotate=90]{Action} (Environment.0);
             \path [line] (Environment.190) --++ (-6em,0em) |- node [anchor=center, above, near start, rotate=90] {New state} (Agent.170);
             \path [line] (Environment.170) --++ (-4.25em,0em) |- node [anchor=center, above, near start, rotate=-90] {Utility} (Agent.190);
        \end{tikzpicture}
        }
    \caption{Independent reinforcement learning.}
    \label{fig:inrl}
\end{wrapfigure}
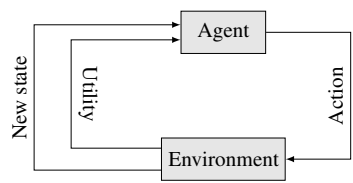{}

The training is done in iterations called \emph{epoch}s. During each epoch, the training algorithm performs two operations on the environment: (1) resetting the environment to the starting state. In return, the environment provides the agent with the initial observation; (2) for each step, the agent decides on an action to take which updates the state of the environment based on transition rules.

Each \emph{epoch} of training is finished and the environment is forced to reset when the number of steps taken in current epoch reaches $T$. It is possible to complete the training without having epochs, however, this condition ensures that 1) the majority of the action/observation space is explored, 2) the training agent is not stuck in a locally optimal state.

Each \emph{step} to the environment updates the state of the system based on the agent's action ($a$) and the current state of the environment ($s$), and returns a new observation ($O$), immediate utility given to agent ($r$), and whether the environment is finished or not. This new information and the previous observation of the agent forms an \emph{experience}. Specifically, an experience is defined as a tuple~of:
\begin{align}
    e = \langle O_\tau, a_\tau, O_{\tau + 1}, r_{\tau} \rangle
\end{align}
where $O_\tau$ and $a_\tau$ are the agent's observation and action at time step $\tau$; and  $O_{\tau+1}$ and $r_{\tau}$ are the agent's observation and immediate utility received at the next time step $\tau+1$. 

The objective of InRL is to find one policy $\pi$ which is a mapping from observation space to action space, such that:
\begin{align}
    &\pi(O_\tau) \mapsto a \label{eq:policy}\\
    \text{which maximizes} \; & U^*_\tau = \mathbb{E}\left[\sum_{t=0}^\infty \gamma^t r_{t + \tau} \middle| \pi\right] \label{eq:policy_objective}
\end{align}
where $\gamma \in [0, 1)$, the \emph{``discount factor,''} prioritizes the rewards claimed at the current time step over the future rewards. When $\gamma = 0$, the player only cares about the current rewards, and when $\gamma = 1$, the player cares about all future rewards equally.

Reinforcement learning aims to maximize the received utility of the agent ($U_*$) by trial and error: interacting with the environment (randomly, heuristics, or referring to the experiences the agent has seen so far), generally, during the training, there are two ways to find actions to be taken at each step: (1) \emph{Exploitation}: we use the currently trained policy to choose actions, which helps the agent to more accurately find $U_*$ values of actions in a state. (2) \emph{Exploration}: to find optimal actions which yields to higher utility by doing random action and exploring the Action/Observation space. One of approaches for deciding on doing exploration or exploitation is the $\epsilon$-greedy where in each step the agent explores with probability $\epsilon$, or take the current optimal action with probability $1-\epsilon$.


\begin{center}
\begin{minipage}[t]{.495\textwidth}
        \vspace{0pt}
        \setlength{\algomargin}{0.8em}
\begin{algorithm}[H]
\SetAlgoLined
\KwResult{policy $\sigma$}
$Q \gets $ random\;
\For{$N_e$ episodes}{
$O \gets $ reset\_game()\;
$\epsilon_\tau \gets $ 1\;
\For{$\tau \in \{0, \ldots, T_{\field{epoch}}$\}}{
\eIf{$random[0, 1] \leq \epsilon_\tau$}{
$a \gets$ random\_action\;
}{
$a \gets \operatorname{argmax}_{a'} Q(S, a')$\;
}
$(S', r) \gets$ step\_game$(a)$\;
add $e = \langle S, S', a, r \rangle$ to $E$\;
sample $X$ from $E$\;
update D$Q$N based on $X$\;
$S \gets S'$\;
decay $\epsilon_\tau$\;
}
}
$\sigma \gets \left\langle S \mapsto \operatorname{argmax}_a Q(S, a) \right\rangle$\;
\caption{InRL}
\label{algo:inrl}
\end{algorithm}
    \end{minipage}%
~\begin{minipage}[t]{.495\textwidth}
        \vspace{0pt}
        \setlength{\algomargin}{0.8em}
\begin{algorithm}[H]
\SetAlgoLined
\KwResult{set of pure policies $\Pi^a$ and $\Pi^d$}
$\Pi^a \gets$ attacker heuristics\;
$\Pi^d \gets$ defener heuristics\;
\While{$U^p(\sigma^p, \sigma^{\bar{p}})$ not converged}{
    $\sigma^a, \sigma^d \gets $ solve\_MSNE($\Pi^a, \Pi^d$)\;
    $\theta \gets$ random\;
    $\pi^a_+ \gets$ train($T \cdot N_e, \field{env}^a[\sigma^d], \theta$)\;
    $\Pi^a \gets \Pi^a \cup \pi^a_+$\;
    assess $\pi^a_+$\;
    $\sigma^a, \sigma^d \gets $ solve\_MSNE($\Pi^a, \Pi^d$)\;
    $\theta \gets$ random\;
    $\pi^d_+ \gets$ train($T \cdot N_e, \field{env}^d[\sigma^a], \theta$)\;
    $\Pi^d \gets \Pi^d \cup \pi^d_+$\;
    assess $\pi^d_+$\;
}
\caption{Adaptive Solver}
\label{algo:do}
\end{algorithm}
    \end{minipage}
\end{center}

\subsection{Deep-$Q$-Network Learning}
\label{subsec:dql}

$Q$-Learning uses a $Q$ function to estimate the expected future utilities of an action in an observation state (Equation~\ref{eq:policy_objective}):
\begin{align}
    Q(O_\tau, a_\tau) = U^*_\tau
\end{align}
Given a tabular approach of storing the $Q$ value for each observation state, we can find the value of the $Q$ function by applying the Bellman optimization equation:
\begin{align}
    Q(O_\tau, a_\tau) = (1-\alpha) &\cdot Q(O_\tau, a_\tau) \label{eq:tdtarget} \\
    + \alpha &\cdot (\underbrace{r_{\tau} + \gamma \cdot \max_{a'}Q(O_{\tau + 1}, a')}_{\text{TD Target}}) \nonumber
\end{align}
where $\alpha$ is the learning rate of the $Q$ function. The idea for updating the $Q$ function is that the $Q$ function should minimize the \textit{temporal difference} (TD) error, \ie the difference between the predicted $Q$ value, and the actual expected utility ($U^*$).

Mnih~\ea~\cite{mnih2013playing} show that it is possible to use \textit{multi layer perceptrons} (MLP) as approximators for the $Q$ function. This is useful since neural networks can generalize the similarities between the observation states and $Q$ values. 
When using the MLP as $Q$ approximator with parameters $\theta$, we define a \textit{mean squared error} (MSE) loss function on a batch $X$ of experiences:
\begin{align}
    L_{\theta} = \frac{1}{|X|}\sum_i^X (q_\tau - Q(O_\tau, a_\tau|\theta))^2 
\end{align}
where $q_\tau$ is the \textit{TD Target} of Equation~\ref{eq:tdtarget} and the optimal action is $\operatorname{argmax}_{a'} Q(O_\tau, a')$: 
\begin{align}
    \label{eq:q_optiomization}
    q_\tau = r_{\tau} + \gamma Q(O_{\tau +1}, \operatorname{argmax}_{a'} Q(O_\tau, a')) | \theta)
\end{align}{}
We can minimize the loss ($L_{\theta}$) with gradient descent and learning rate $\alpha$.

\section{Problem Formulation}
\label{sec:problem}

In Section~\ref{sec:model}, we build an MTD model using a MAPOMDP. In this section, based on this model, we design an adversarial game between the adversary (denoted by $p = 1$ or $a$) and the defender (denoted by $p = 0$ or $d$). In this setting, we assume that each player chooses a strategy to play, where each strategy is a policy function that given the current observation of the environment, returns an action to be taken. 
As we assumed that each playing strategy is a policy function, in this paper, we use the terms ``strategy'' and ``policy'' interchangeably.

\subsection{Pure Strategy}
\label{subsec:pure}

A pure strategy $\pi^p$ for player $p$ is a deterministic policy function $\pi^p(O^p) \mapsto a^p$ which given player $p$'s current observation of the system ($O^p$) produces an action $a^p$ to be taken by this player. We denote the set of pure strategy space for player $p$ in this adversarial game as $\Pi^p$. 

When players are following pure policies $\pi^a \in \Pi^a$ and $\pi^d \in \Pi^d$, their expected utility can be expressed as sum of discounted future rewards with discount factor of $\gamma$. Formally:
\begin{align}
    \forall_{p \in \{1, 0\}} :
    U^p(\pi^{p}, \pi^{\bar{p}}) = \mathbb{E}\left[ \sum_{t=0}^{\infty} \gamma^t \cdot r^p_{t} \middle| \pi^{p}, \pi^{\bar{p}} \right]
\end{align}

\subsection{Mixed Strategy}
\label{subsec:mixed}

One way to express stochastic policies is to use probability distributions over pure policies. A mixed strategy of player $p$ is a probability distribution $\sigma^p = \{\sigma^p(\pi^p)\}_{\pi^p \in \Pi^p}$ over the player's pure strategies $\Pi^p$ where $\sigma^p(\pi^p)$ is the probability that player $p$, chooses policy $\pi^p$. 


We denote $\Sigma^p$ as the strategy space of player $p$. The utility profile of the adversary and the defender when they are following mixed strategies $\sigma^{a} \in \Sigma^{a}$ and $\sigma^{d} \in \Sigma^{d}$, respectively, can be calculated as:
\begin{align}
    &\forall_{p \in \{1, 0\}} :
    &U^p(\sigma^{p}, \sigma^{\bar{p}}) = \sum_{\pi^{p} \in \Pi^{p}} \sum_{\pi^{\bar{p}} \in \Pi^{\bar{p}}} \sigma^{p}(\pi^{p}) \cdot \sigma^{\bar{p}}(\pi^{\bar{p}}) \cdot U^p(\pi^{p}, \pi^{\bar{p}})
\end{align}

Note that the notation of utility profile of players when they follow pure strategies is overloaded to also support the use of mixed strategies.

\subsection{Solution Concept}
\label{subsec:solution}

The aim of both players is to maximize their utility. As we are considering a strong adversary and defender, we can assume that they always pick the strategy which maximizes their own utility. A \emph{``best response''} mixed strategy $\sigma_*^p(\sigma^{\bar{p}})$ ensures maximum utility for the player choosing it ($p$) while the opponent is using mixed strategy $\sigma^{\bar{p}}$. In other words, the agent choosing a best response mixed strategy $\sigma_*^p$ cannot gain more utility without having the opponent changing its strategy. When the player $p$ is using a mixed strategy $\sigma^P$, the opponent's best response is:
\begin{align}
    \sigma_*^p(\sigma^{\bar{p}}) &= \operatorname{argmax}_{\sigma^p} U^p(\sigma^{p}, \sigma^{\bar{p}})
\end{align}

We optimize one player's strategy assuming that the opponent always uses a best response strategy to the player's strategy, $\sigma^p = \sigma^p_*(\sigma^{\bar{p}})$. 
This formulation of a general-sum game is in fact equivalent to finding a \textit{mixed-strategy Nash equilibrium} (MSNE) of players' policy space $\Pi^{a}$ and $\Pi^{d}$. In other words, a combination of strategies $(\sigma_*^p, \sigma_*^{\bar{p}})$ is MSNE, \emph{iff}:
\begin{align}
    \forall_{p \in \{1, 0\}} \forall_{\sigma^p \in \Sigma^p}:
    U^p(\sigma_*^p, \sigma_*^{\bar{p}}) \geq U^a(\sigma^p, \sigma_*^{\bar{p}}) \label{eq:nash_eq}
\end{align}

That is, neither agent can increase its expected utility without having the opponent changing its strategy. Therefore, our solution must find the MSNE of the MTD game where $\Pi^{a}$ and $\Pi^{d}$ are the policy space of the players.

\section{Framework}
\label{sec:framework}

In Section~\ref{sec:model}, we described a moving target defense model as a \textit{Multi-Agent Partially- Observable Markov Decision Process} (MAPOMDP). In Section~\ref{sec:problem}, we proposed a game based on this model and concluded that finding an optimal action policy for the adversary and the defender is equivalent to finding the MSNE of the game. In this section, we propose a computational approach and build a framework atop of the double oracle (Section~\ref{subsec:do}) and D$Q$L (Section~\ref{subsec:dql}) algorithms to find the optimal action policies for the adversary and the defender.

\subsection{Solution Overview}
\label{subsec:do}

The iterative \emph{Double Oracle} (DO) algorithm\cite{mcmahan2003planning}, solves the MSNE of a game given an arbitrary subset of policy space for each player ($\Pi^p_0 \subset \Pi^p$). We denote the explored policy space of player $p$ until iteration $\tau$ of the DO algorithm by $\Pi^p_\tau$. In each iteration $\tau$ of the DO algorithm, for each player, a pure strategy best response ($\pi^*$) to the MSNE of the opponent is calculated using a \emph{``best response oracle''} and added to the strategy sets. Formally, in each iteration:
\begin{align}
    \forall_{p \in \{1, 0\}}:
    \Pi^p_{\tau + 1} \gets \Pi^p_\tau \cup \{\pi^p_*(\sigma^{\bar{p}}_{*, \tau})\}
\end{align}
where $\sigma^p_{*, \tau}$ is the MSNE of the player $p$ given the strategy sets $\Pi^p_\tau$. The DO algorithm guarantees~\cite{mcmahan2003planning} the convergence of the MSNE of these strategy sets to the actual MSNE of the game as long as the policy space for both players are finite. However, as the policy space of the players are huge (still finite), enumeration of the policy space in search of the best response is infeasible.

For each player, we can use a tabular based InRL such as $Q$-Learning\cite{watkins1992q} algorithm as a best response oracle to find a best response pure strategy to the opponent's MSNE. Since the opponent's strategy is fixed, the agent learning through reinforcement learning can treat the opponent's actions as part of its localized environment without overfitting to the opponent's policies. 

\subsection{Challenges}
\label{subsec:challenges}

Solving the MAPOMDP model of Section~\ref{sec:model} with the DO algorithm is not straightforward. In the following subsections, we discuss the issues faced while solving the MAPOMDP model and propose approaches for resolving these issues.

\subsubsection{Partial Observability}
\label{subsec:partial}
For both players, state is only partially observable.
This can pose a significant challenge for the defender, who does not even know whether a server is compromised or not. 
Consider, for example, the defender observing that a particular server has been probed only a few times: this may mean that the server is safe since it has not been probed enough times; but it may also mean that the adversary is not probing it because the server is already compromised.
We can try to address this limitation by allowing the defender's policy to consider a long history of preceding states; however, this poses computational challenges since the size of the effective state space for the policy explodes.

Since partial observability poses a challenge for the defender, we let the defender's policy use information from preceding states.
To avoid state-space explosion, we feed this information into the policy in a compact form.
In particular, we extend the observed state of each server (i.e., number of observed probes and whether the server is online) with (a) the amount of time since the last reimaging and (b) the amount of time since the last observed probe. So, the actual state presentation of the defender will be:
\begin{align}
    O_i^d = \langle &\field{status},
    \field{time\_to\_up},
    \field{progress}, \nonumber \\ 
    &\field{time\_since\_last\_probe}, \nonumber \\
    &\field{time\_since\_last\_reimage} \rangle
\end{align}

Similarly, the adversary should probe the servers which where probed in many steps in the past to make sure that its progress was not reset. So, we add the amount of time since the last probe to its observation state.
\begin{align}
    O_i^a = \langle &\field{status}, 
    \field{time\_to\_up}, 
    \field{progress}, \field{control}, \nonumber \\
    &\field{time\_since\_last\_probe} \rangle 
\end{align}

\subsubsection{Complexity of MSNE Computation}

In zero-sum games, computation of MSNE can be done using linear programming which has a polynomial time complexity. However, in general-sum games, the solving algorithms is PPAD-complete\cite{shoham2008multiagent} which makes it infeasible for solving a game of non-trivial size. Therefore, we use an \emph{``$\epsilon$-equilibrium''} solver, which produces an approximate correct result. One such solver is the Global Newton solver\cite{govindan2003global}.

\subsubsection{Equilibrium Selection}
Typically, the DO algorithm is used with zero-sum games which all the equilibria of the game yield the same payoff for each player. However, in general-sum games, there may exist multiple equilibria with significantly different payoffs. The DO algorithm in general-sum games only converges to one of these equilibria. The exact equilibrium the DO algorithm converges to depends on the initial policy space of the players and the exact output of the best response oracle. However, in our experiments (Section~\ref{subsec:results}), we show that in practice, in adversarial games, this problem is not significant, \ie all equilibria produce almost the same results (Table~\ref{tab:env_eq}) indeferent to the initial policy space.

\subsubsection{Model Complexity}
\label{subsec:complexity}

Due to the complexity of our MAPOMDP model, computation of best response using tabular InRL approaches is computationally infeasible. In fact, even representation of a single policy as keeping the best action for all the observation states is infeasible. Further, tabular approaches fail to generalize the relations between observations and actions. Thus, the action/observation space need to be enumerated many times in order for the algorithm to produce correct results.

To address this challenge, we can use computationally feasible \emph{``approximate best responses''} to produce an approximate best response pure strategy($\pi_+$) instead of true best responses. Lanctot~\ea~\cite{lanctot2017unified} show that deep reinforcement learning can be used as an approximate best response oracle. However, when approximate best responses are used instead of true best responses, convergence guarantees are lost. In our experiments, we show that this algorithm indeed converges with six iterations (see Figure~\ref{fig:do_iter}). 

\subsubsection{Short-term Losses vs. Long-term Rewards}
\label{subsec:gamma}
For both players, taking an action has a negative short-term impact:
for the defender, reimaging a server results in lower rewards while the server is offline; for the adversary, probing incurs a cost.
While these actions can have positive long-term impact, benefits may not be experienced until much later:
for the defender, a reimaged server remains offline for a long period of time;
for the attacker, many probes may be needed until a server is finally compromised.

As a result, with typical temporal discount factors (e.g., $\gamma = 0.9$), it may be an optimal policy for a player to never take any action since the short-term negative impact outweighs the long-term benefit.
In light of this, we can use higher temporal discount factors (e.g., $\gamma = 0.99$).
However, such values can pose challenges for deep reinforcement learning since it will be much more difficult to converge.

\subsection{Solution Approach}
\label{subsec:approach}

Prakash and Wellman~\cite{prakash2015empirical} proposed a set of heuristic strategies for each player (described in Section~\ref{subsec:heuristics}). However, as these strategies are only a subset of the agents' policy spaces, their MSNE is not necessarily the MSNE of the $\Sigma^a$ and $\Sigma^d$. In Section~\ref{subsec:do}, we show how we can find the MSNE of the game, given a subset of policy space for each agent. In this section, we propose our framework to find the MSNE of the MTD game and therefore, the optimal decision making policy for the adversary and the defender. Algorithm~\ref{algo:do} shows a pseudo-code of our framework.

We start by initializing the adversary's and defender's strategy sets with heuristic policies (Section~\ref{subsec:heuristics}). From this stage, we proceed in iterations. In each iteration, first, we compute MSNE of the game restricted to the current strategy sets $\Pi^a$ and $\Pi^d$, and take the adversary's equilibrium mixed strategy $\sigma^a$ and train an approximate best-response policy ($\pi_+^d(\sigma^a)$) for the defender assuming that the adversary uses $\sigma^a$. Next, we add this new policy to the set of policies of the defender ($\Pi^d \gets \Pi^d \cup \{ \pi_+^d \}$).

Then, we do the same for the adversary. First, find the MSNE strategy of the defender ($\sigma^d$), and train an approximate best-response policy ($\pi_+^a(\sigma^d)$) for the adversary assuming that the defender uses $\sigma^d$. Then, we add this new policy to set of policies for the adversary ($\Pi^a \gets \Pi^a \cup \{ \pi_+^a \}$).

In both cases, when computing an approximate best response ($\pi_+(\sigma_*)$) for a player against its opponent's MSNE strategy $\sigma_*$, the opponent's strategy $\sigma_*$ is fixed, so we may consider it to be part of the player's environment. As a result, we can cast the problem of finding an approximate best response for agent as \textit{Independent Reinforcement Learning} (InRL). Each iteration of InRL, defined as \texttt{train()} in Algorithm~\ref{algo:do}, receives a total number of steps $T$ of training, and initial parameters $\theta$. Moreover, we denote the InRL environment for the player $p$ when the opponent plays with a mixed strategy $\sigma^{\bar{p}}$ as $\field{env}^p[\sigma^{\bar{p}}]$.

As we are dealing with discrete action/observation spaces in the MTD model, D$Q$L \cite{mnih2013playing} is a suitable InRL algorithm for finding an approximate best response. In each time step of the InRL, both players need to decide on an action. The learning agent either chooses an action randomly (\ie exploration), or follows its current policy. The opponent whose strategy is fixed to a mixed strategy $\sigma^{\bar{p}}$ refers to a pure strategy $\pi^{\bar{p}} \in \Pi^{\bar{p}}i$ with probability distribution $\sigma^{\bar{p}}$, and follows that policy.

The MSNE payoff evolves over these iterations whenever we add a new policy for an agent, which is trained against the best mixed strategy of the opponent, the MSNE changes in favor of the agent. We continue these iterations until the MSNE payoff of the defender and the adversary ($U^p(\sigma^p_*, \sigma^{\bar{p}}_*)$) converges. Formally, we say that the for both players the MSNE is converged \textit{iff}:
\begin{align}
    \forall_{p \in \{1, 0\}}:
    U^p(\pi^p_+, \sigma^{\bar{p}}) \leq U^p(\sigma^p, \sigma^{\bar{p}})
\end{align}
where $(\sigma^p)$ is the previous MSNE and $\pi^p_+$ is the approximate best response found for player $p$ in opposing to the opponent's MSNE at the current iteration. This convergence means that neither the adversary nor defender could perform better by introducing new policies.

\section{Evaluation}
\label{sec:eval}

In this section, first we describe the heuristic strategies of the MTD game (Section~\ref{subsec:heuristics}). Next, we describe our implementation of the framework (Section~\ref{subsec:impl}). Finally, we discuss the numerical results (Section~\ref{subsec:results}).

\subsection{Baseline Heuristic Strategies}
\label{subsec:heuristics}



Prakash and Wellman~\cite{prakash2015empirical} proposed a set of heuristic strategies for both the adversary and the defender. Earlier, we used these strategies as our initial policy space for the DO algorithm. In this section, we describe these heuristics.

\subsubsection{Adversary's Heuristic Strategies}

\begin{itemize}
    \item \emph{Uniform-Uncompromised:} Adversary launches a probe every~$P_A$ time steps, always selecting the target server uniformly at random from the servers under the defender's control.
    \item \emph{MaxProbe-Uncompromised:} Adversary launches a probe every~$P_A$ time steps, always targeting the server under the defender's control that has been probed the most since the last reimage (breaking ties uniformly at random).
    \item \emph{Control-Threshold:} Adversary launches a probe if the adversary controls less than a threshold~$\tau$ fraction of the servers, always targeting the server under the defender's control that has been probed the most since the last reimage (breaking ties uniformly at random).
    \item \emph{No-Op:} Adversary never launches a probe.
\end{itemize}

\subsubsection{Defender's Heuristic Strategies}

\begin{itemize}
    \item \emph{Uniform:} Defender reimages a server every~$P_D$ time steps, always selecting a server that is up uniformly at random.
    \item \emph{MaxProbe:} Defender reimages a server every~$P_D$ time steps, always selecting the server that has been probed the most (as observed by the defender) since the last reimage (breaking ties uniformly at random).
    \item \emph{Probe-Count-or-Period (PCP):} Defender reimages a server which has not been probed in the last $P$ time steps or has been probed more than $\pi$ times (selecting uniformly at random if there are multiple such servers).
    \item \emph{Control-Threshold:} Defender assumes that all of the observed probes on a server except the last one were unsuccessful. Then, it calculates the probability of a server being compromised by the last probe as $1-e^{-\alpha \cdot (\rho + 1)}$. Finally, if the expected number of servers in its control is below $\tau \cdot M$ and it has not reimaged any servers in $P_D$, then it reimages the server with the highest probability of being compromised (breaking ties uniformly at random). In other words, it reimages a server \emph{iff}
    \begin{align}
        \mathbb{E}[n_c^d] \leq M\cdot\tau 
    \end{align}
    and the last reimage was at least $P_D$ time steps ago.
    
    
    \item \emph{No-Op:} Defender never reimages any servers.
\end{itemize}

Table~\ref{tab:util} shows the expected rewards for all combinations of heuristic defender and adversary strategies 
in an environment whose parameters 
are described in Table~\ref{tab:symbols}. Also, in this table, we compare our mixed strategy policies computed by D$Q$L to these strategies.

\begin{table*}
\small
    \caption{Payoff Table for Heuristic Adversary vs. Defender. For comparison, MSNE strategy payoffs are shown.}
    \label{tab:util}
    \centering
    \resizebox{\linewidth}{!}{
    \begin{tabular}{|c|c|c|c|c|c||c|}\hline

\backslashbox{Adversary}{Defender} & No-OP & ControlThreshold & PCP & Uniform & MaxProbe & Mixed Strategy D$Q$L \\
\hline

No-OP & \backslashbox{26.89}{98.20} & \backslashbox{26.89}{98.20} & \backslashbox{26.89}{98.20} & \backslashbox{46.03}{95.83} & \backslashbox{26.89}{98.20} & \backslashbox{33.23}{97.47} \\
\hline
MaxProbe & \backslashbox{78.66}{47.69} & \backslashbox{75.67}{49.62} & \backslashbox{36.58}{93.01} & \backslashbox{64.56}{67.12} & \backslashbox{41.99}{86.82} & \backslashbox{45.87}{87.84}  \\
\hline
Uniform & \backslashbox{79.08}{46.74} & \backslashbox{70.97}{51.58} & \backslashbox{44.43}{89.48} & \backslashbox{56.83}{76.23} & \backslashbox{57.14}{75.21} & \backslashbox{45.91}{88.16}  \\
\hline
ControlThreshold & \backslashbox{63.64}{85.98} & \backslashbox{65.58}{85.35} & \backslashbox{46.38}{88.81} & \backslashbox{59.54}{81.32} & \backslashbox{60.43}{80.09} & \backslashbox{45.91}{87.91}  \\
\hline  
\hline
Mixed Strategy D$Q$L & \backslashbox{62.78}{72.29} & \backslashbox{58.31}{82.45} & \backslashbox{45.76}{91.32} & \backslashbox{55.31}{87.10} & \backslashbox{44.57}{91.32} & \backslashbox{45.23}{92.38}  \\
\hline
    \end{tabular}
    }
\end{table*}

\subsection{Implementation}
\label{subsec:impl}

We implemented the MAPOMDP of Section~\ref{sec:model} as an Open AI Gym~\cite{gym} environment. We used Stable-Baselines' DQN\cite{stable-baselines} as the implementation of the D$Q$L. Stable-Baselines internally uses TensorFlow~\cite{abadi2016tensorflow} as the neural network framework. For the artificial neural network as our $Q$ approximator, we used a feed forward network with two hidden layers of size 32, and \emph{tanh} as our activation function. The rest of parameters are available at Table~\ref{tab:symbols}. We implemented the remainder of our framework in Python, including the double oracle algorithm. For computation of the mixed strategy $\epsilon$-equilibrium of a general-sum game, we used the Gambit-Project's Global Newton implementation \cite{mckelvey2006gambit}.

We run the experiments on a computer cluster where each node has two 14 Cores 2.4 GHz Intel Xeon CPU and two NVIDIA P100 GPU. Each node is capable of running $\approx 75$ steps of D$Q$L per second, which yields to 1.5 hours per each reference to the best response oracle (\ie D$Q$L training for an adversary or a defender). Further, the D$Q$L algorithm is not distributed, so we only use one core of the CPU. This paves the way for multiple D$Q$Ls to run at the same time. We also need to mention that, set of policies only needs to be pre-computed. While policies are in use, inference takes only milliseconds.

\subsection{Numerical Results}
\label{subsec:results}

\begin{figure}
    \centering
    \begin{subfigure}[t]{.49\linewidth}
    \heavyFigure{
    \pgfplotstableread[col sep=comma]{data/attacker_curve.csv}\attacker
    \pgfplotstableread[col sep=comma]{data/defender_curve.csv}\defender
    \begin{tikzpicture}
        \begin{axis}
        [
            ylabel=Utility,
            xlabel=Step,
            legend pos=south east,
            width=\linewidth,
            height=\linewidth,
            legend style={nodes={scale=0.8, transform shape}}
        ]
            \addplot [blue, smooth] table [x=Step, y=Value] {\attacker};
            \addlegendentry{Attacker}
            \addplot [red, smooth] table [x=Step, y=Value] {\defender};
            \addlegendentry{Defender}
        \end{axis}
    \end{tikzpicture}}
    \caption{Learning curve of the agents}
    \label{fig:learning_curve}
\end{subfigure}
    \hfill
    \begin{subfigure}[t]{.49\linewidth}
    \centering
    \heavyFigure{
    \pgfplotstableread[col sep=comma]{data/do_curve.csv}\do
    \begin{tikzpicture}
        \begin{axis}
        [
            ylabel=Utility,
            xlabel=Iteration,
            width=\linewidth,
            height=\linewidth,
        ]
            \addplot [red, smooth, mark=square] table [x expr=\coordindex, y=attacker-0] {\do};
            \addplot [blue, smooth, mark=square] table [x expr=\coordindex, y=defender-0] {\do};
            
            \addplot [red, smooth, mark=star] table [x expr=\coordindex, y=attacker-1] {\do};
            \addplot [blue, smooth, mark=star] table [x expr=\coordindex, y=defender-1] {\do};
            
            \addplot [red, smooth, mark=x] table [x expr=\coordindex, y=attacker-2] {\do};
            \addplot [blue, smooth, mark=x] table [x expr=\coordindex, y=defender-2] {\do};
        \end{axis}
    \end{tikzpicture}}
    \caption{Evolution of payoff}
    \label{fig:do_iter}
\end{subfigure}
    \caption{In Figure~\ref{fig:do_iter}, iteration 0 shows the MSNE payoff of the heuristics while each D$Q$N training for adversary and defender happens at odd and even iterations, respectively.}
\end{figure}
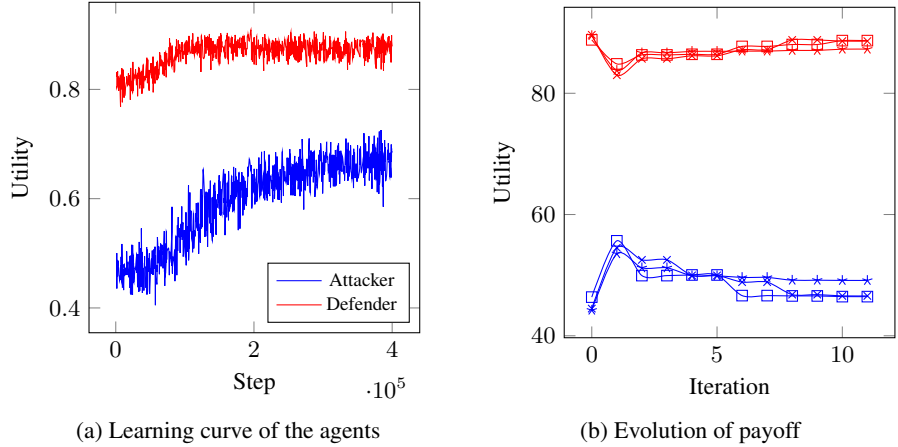

Figure~\ref{fig:learning_curve} shows the learning curve of the agents for their first iteration of the DO algorithm (Iteration 1 and 2), with the MTD environment whose parameters are specified in Table~\ref{tab:symbols}. We can see the improvement of the resulted policy compared to the heuristics (Table~\ref{tab:util}).

Figure~\ref{fig:do_iter} shows the evolution of MSNE payoff over the iterations of the DO algorithm with environment parameters of Table~\ref{tab:symbols}. In this figure, each training for the adversary and defender happens at odd and even iterations, respectively, while iteration 0, is the payoff equilibrium of heuristic policies. Also, this figure shows that the DO algorithm indeed converges with $\approx$ 4 trainings for each player, \ie 6 hours of training in total. Comparing multiple runs with the same configuration (highlighted runs in Table~\ref{tab:env_eq}), shows that the DO algorithm with multiple approximations (\eg approximation with deep networks, approximation on equilibrium computation). 

The converged equilibrium payoffs for the adversary and the defender for different environment parameters are shown in Table~\ref{tab:env_eq}. These payoffs are in fact in agreement with \emph{empirical game theoretic analysis} EGTA done by Prakash and Wellman~\cite{prakash2015empirical}. 

To analyse the impact of equilibrium selection on the MSNE payoff of the game, we executed Algorithm~\ref{algo:do} with the same environment parameters of gray rows of Table~\ref{tab:env_eq}, but without heuristics as initial policies. The initial policies are set to NoOP adversary and NoOP defender. As we can see in Table~\ref{tab:env_eq}, the payoffs for the adversary and defender are almost identical in both cases (using heuristics as initial policy space and NoOP as initial policy space for players).

\section{Related Work}
\label{sec:related}

\begin{wraptable}{r}{.4\linewidth}
\centering
\caption{Final Equilibrium Payoff of MTD environment with different configurations}
\label{tab:env_eq}
\resizebox{\linewidth}{!}{
    \begin{tabular}{|l|l|r|r|l|l|l|r|r|}
\hline
$M$ & utenv & $\theta^p_{th}$ & \multicolumn{1}{l|}{$C_A$} & $\Delta$ & $\alpha$ & $\nu$ & \multicolumn{1}{l|}{Adversary} & \multicolumn{1}{l|}{Defender} \\ \hline
\multicolumn{9}{|c|}{Heuristics as Initial Policy Space} \\ \hline
\rowcolor{Gray} 10  & 2     & 0.2       & 0.2                        & 7        & 0.05     & 0     & 46.46                          & 88.69                         \\ \hline
\rowcolor{Gray} 10  & 2     & 0.2       & 0.2                        & 7        & 0.05     & 0     & 49.13                          & 87.27                         \\ \hline
10  & 2     & 0.2       & 0.05                       & 7        & 0.05     & 0     & 63.26                          & 86.31                         \\ \hline
10  & 2     & 0.2       & 0.1                        & 7        & 0.05     & 0     & 57.83                          & 87.21                         \\ \hline
10  & 2     & 0.2       & 0.05                       & 7        & 0.05     & 0     & 63.03                          & 87.37                         \\ \hline
10  & 2     & 0.2       & 0.1                        & 7        & 0.05     & 0     & 58.08                          & 86.55                         \\ \hline
\rowcolor{Gray} 10  & 2     & 0.2       & 0.2                        & 7        & 0.05     & 0     & 46.58                          & 88.60                         \\ \hline
10  & 0     & 0.2       & 0.05                       & 7        & 0.05     & 0     & 35.05                          & 92.07                         \\ \hline
10  & 0     & 0.5       & 0.05                       & 7        & 0.05     & 0     & 8.98                           & 77.27                         \\ \hline
10  & 0     & 0.8       & 0.05                       & 7        & 0.05     & 0     & 1.80                           & 73.10                         \\ \hline
10  & 0     & 0.2       & 0.1                        & 7        & 0.05     & 0     & 30.87                          & 90.81                         \\ \hline
10  & 0     & 0.5       & 0.1                        & 7        & 0.05     & 0     & 7.59                           & 92.41                         \\ \hline
10  & 0     & 0.8       & 0.1                        & 7        & 0.05     & 0     & 1.80                           & 73.10                         \\ \hline
10  & 0     & 0.2       & 0.2                        & 7        & 0.05     & 0     & 26.89                          & 98.20                         \\ \hline
10  & 0     & 0.2       & 0.1                        & 3        & 0.05     & 0     & 30.64                          & 94.46                         \\ \hline
10  & 1     & 0.2       & 0.1                        & 7        & 0.05     & 0     & 26.89                          & 98.20                         \\ \hline
10  & 1     & 0.2       & 0.05                       & 7        & 0.05     & 0     & 26.89                          & 98.20                         \\ \hline
10  & 1     & 0.2       & 0.2                        & 7        & 0.05     & 0     & 26.89                          & 98.20                         \\ \hline
10  & 1     & 0.2       & 0.1                        & 3        & 0.05     & 0     & 26.89                          & 98.20                         \\ \hline
10  & 3     & 0.2       & 0.1                        & 7        & 0.05     & 0     & 74.16                          & 97.83                         \\ \hline
10  & 3     & 0.2       & 0.05                       & 7        & 0.05     & 0     & 79.31                          & 97.90                         \\ \hline
10  & 3     & 0.2       & 0.2                        & 7        & 0.05     & 0     & 65.21                          & 97.66                         \\ \hline
\multicolumn{9}{|c|}{NoOP as Initial Policy Space} \\ \hline
10 & 2 & 0.2 & 0.2 & 7 & 0.05 & 0 & 50.72 & 90.23  \\ \hline
10 & 2 & 0.2 & 0.2 & 7 & 0.05 & 0 & 46.34 & 89.83 \\ \hline
10 & 2 & 0.2 & 0.2 & 7 & 0.05 & 0 & 47.68 & 88.92 \\ \hline
\end{tabular}
}
\end{wraptable}

In this work, we used multi agent reinforcement learning to find optimal policies for the adversary and the defender in an MTD game model.
In prior work, researchers have investigated both the application of reinforcement learning in cyber-security (Section~\ref{sec:relatedRL}) and game-theoretic models for MTD (Section~\ref{sec:relatedMTD}). 
Perhaps the most closely related work on integration of reinforcement learning and moving target defense is the work done by Sengupta and Kambhampati\cite{sengupta2020multi}. They propose a Bayesian Stackelberg game model to MTD and solve (\ie finding the optimal action policy for the defender) it using $Q$-Learning. However, their approach is not applicable in our model since 1) Our model is not a Stackleberg game; neither the adversary nor the defender observes the actions taken by the opponent, and 2) Their proposed model has less complexity, making table based $Q$-Learning feasible.

\subsection{Moving Target Defense}
\label{sec:relatedMTD}

One of the main research areas in moving target defense is to model interactions between the adversaries and the defenders. In the area of game-theoretic models for moving target defense, the most closely related work is from Prakash~\ea~\cite{prakash2015empirical}, which introduces the model that our work uses. This model can also be used for defense against DDoS attacks\cite{wright2016moving}, and defense for web applications\cite{sengupta2017game}. Further, in this area, researchers have proposed MTD game models based on Stackelberg games\cite{li2019optimal}, Markov Games\cite{lei2017optimal,tan2019optimal}, Markov Decision Process\cite{zheng2019markov}, and FlipIt game \cite{oakley2019playing}.

For solving a game model (\ie finding the optimal playing strategies), numerous approaches such as solving a min-max problem~\cite{li2019optimal}, non-linear programming~\cite{lei2017optimal}, Bellman equation~\cite{tan2019optimal,zheng2019markov}, Bayesian belief networks\cite{albanese2019moving}, and reinforcement learning\cite{eghtesad2019deep,hu2019reinforcement,oakley2019playing} has been suggested.

\subsection{Reinforcement Learning for Cyber Security}
\label{sec:relatedRL}

Usage of machine learning and especially \textit{deep reinforcement learning} (DRL) for cyber security has gained attention recently. Nguyen~\ea~\cite{nguyen2019deep} surveyed current literature on applications of DRL on cyber security. These applications include: DRL-based security methods for cyber-physical systems, autonomous intrusion detection techniques, and multi-agent DRL-based game theory simulations for defense strategies against cyber attacks.

For example, Malialis~\cite{malialis2015distributed,malialis2015distributed2} applied multi-agent deep reinforcement learning on network routers to throttle the processing rate in order to prevent \textit{distributed denial of service} (DDoS) attacks. Bhosale~\ea~\cite{bhosale2014cooperative} proposed a cooperative multi-agent reinforcement learning for intelligent systems~\cite{herrero2009multiagent} to enable quick responses. Another example for multi-agent reinforcement learning is the fuzzy $Q$-Learning approach for detecting and preventing intrusions in \textit{wireless sensor networks} (WSN) by Shamshirband~\ea~\cite{shamshirband2014cooperative}. Furthermore, Tong~\ea~\cite{tong2019finding} proposed a multi-agent reinforcement learning framework for alert correlation based on double oracles.

Iannucci~\ea~\cite{iannucci2019performance} use deep reinforcement learning to find the optimal policy in intrusion response systems. They evaluate the performance of their algorithm based on different configurations of the model, showing that it can be much faster than traditional $Q$-learning.

\section{Conclusion}
\label{sec:conclusion}

Moving target defense tries to increase adversary's uncertainty and attack cost by dynamically changing host and network configurations. In this paper, we have proposed a multi-agent reinforcement learning approach for finding MTD strategies based on an adaptive MTD model. For improvement of the performance of agents in partially observable environments, we proposed a compact memory presentation for the agents. Further, we show that the double oracle algorithm with D$Q$L as best-response oracle is a feasible and promising solution for finding the optimal actions in general-sum adversarial games as it is stable.

\bibliographystyle{splncs04}
\bibliography{main}

\end{document}